\documentclass[12pt]{article}
\def\option{\noindent}
\def\vs{\vspace}
\def\Tr{{\rm Tr}}
\def\mod{{\rm \;mod\;}}
\def\forodd{{\rm \;for\;} N+K {\rm \; odd}}
\def\gcd{{\rm  gcd}}
\def\lcm{{\rm  lcm}}
\def\e{{\rm e}}
\def\ie{{\it i.e.}}
\def\viz{{\it viz.}}
\def\eg{{\it e.g.}}
\def\lam{\lambda}
\def\sig{\sigma}
\def\Dr{{ \Delta }}
\def\Lam{\Lambda}
\def\boundary{  | B \rangle\!\rangle }
\def\ishilam{  | \lam \rangle\!\rangle_I }
\def\ishimu{  | \mu \rangle\!\rangle_I }
\def\cardy{  | \lam \rangle\!\rangle_C }
\def\dualcardy{  | \tlam \rangle\!\rangle_C }
\def\sigcardy{  | \sigma(\lam) \rangle\!\rangle_C }
\def\idcardy{  | \id \rangle\!\rangle_C }

\def\id{ 0 } 
\def\Pplus{ P_+^K }
\def\Smat{ S_{\mu\nu} }
\def\tSmat{ \tS_{\tmu\tnu} }
\def\Nabc{ {N_{\mu\nu}}^{\lam}  }
\def\bNabc{  { {\overline N}_{\mu\nu}     }^{\lam}    }
\def\LR{   {{\overline N}_{\lam'\lam''}}^{\lam} }
\def\tNabc{ {{\tN}_{\tmu \tnu}}^{~~\sig^\Dr(\tlam) } }
\def\cH{  {\cal H}  }
\def\tlam{{\tilde{\lambda}}}
\def\hlam{{\hat{\lambda}}}
\def\tLam{{\tilde{\Lambda}}}
\def\tmu{{\tilde{\mu}}}
\def\tnu{{\tilde{\nu}}}
\def\tN{{\tilde{N}}}
\def\tQ{{\tilde{Q}}}
\def\tS{{\tilde{S}}}
\def\bz{{\bar z}}

\def\bV{\overline{V}}
\def\bT{\overline{T}}
\def\bJ{\overline{J}}
\def\g{g}
\def\sun{{{\rm su}(N)}}
\def\suk{{{\rm su}(K)}}
\def\sup{{{\rm su}(N+K)}}
\def\sumo{{{\rm su}(M_1)}}
\def\sumt{{{\rm su}(M_2)}}
\def\sump{{{\rm su}(M_1+M_2)}}
\def\suminusk{{{\rm su}(-K)}}
\def\hgk{{\hat{g}_K}}   
\def\sunk{{\widehat{\rm su}(N)_K}}
\def\sukn{{\widehat{\rm su}(K)_N}}
\def\one{  {\vcenter  {\vbox  
              {\hrule height.4pt
               \hbox {\vrule width.4pt  height3pt  
                      \kern3pt 
                      \vrule width.4pt  height3pt }
               \hrule height.4pt}
                         }
                   }
           }
\def\two{  {\vcenter  {\vbox  
              {\hrule height.4pt
               \hbox {\vrule width.4pt  height3pt  
                      \kern3pt 
                      \vrule width.4pt  height3pt 
                      \kern3pt
                      \vrule width.4pt height3pt}
               \hrule height.4pt}
                         }
              }
           }
\def\four{  {\vcenter  {\vbox  
              {\hrule height.4pt
               \hbox {\vrule width.4pt  height3pt  
                      \kern3pt 
                      \vrule width.4pt  height3pt 
                      \kern3pt
                      \vrule width.4pt  height3pt 
                      \kern3pt
                      \vrule width.4pt  height3pt 
                      \kern3pt
                      \vrule width.4pt height3pt}
               \hrule height.4pt}
                         }
              }
           }
\def\oneone{ {\vcenter  {\vbox  
              {\hrule height.4pt
               \hbox {\vrule width.4pt  height3pt  
                      \kern3pt 
                      \vrule width.4pt  height3pt }
               \hrule height.4pt
               \hbox {\vrule width.4pt  height3pt  
                      \kern3pt 
                      \vrule width.4pt  height3pt }
               \hrule height.4pt}
                         }
              }
           }
\def\oneoneoneone{ 
              {\vcenter  {\vbox  
              {\hrule height.4pt
               \hbox {\vrule width.4pt  height3pt  
                      \kern3pt 
                      \vrule width.4pt  height3pt }
               \hrule height.4pt
               \hbox {\vrule width.4pt  height3pt  
                      \kern3pt 
                      \vrule width.4pt  height3pt }
               \hrule height.4pt
               \hbox {\vrule width.4pt  height3pt  
                      \kern3pt 
                      \vrule width.4pt  height3pt }
               \hrule height.4pt
               \hbox {\vrule width.4pt  height3pt  
                      \kern3pt 
                      \vrule width.4pt  height3pt }
               \hrule height.4pt}
                         }
              }
           }
\def\be{\begin{equation}}
\def\ee{\end{equation}}
\def\bea{\begin{eqnarray}}
\def\eea{\end{eqnarray}}

\newcommand{\R}{\mathrm{I}\kern -2.5pt \mathrm{R}}
\newcommand{\Z}{\mathsf{Z}\kern -5pt \mathsf{Z}}
\newcommand{\1}{{1\kern -3pt \mathrm{l}}}

\def\theequation{\thesection.\arabic{equation}}

\begin{document}
\bibliographystyle{bst}

\begin{flushright}
{\tt hep-th/0511083}\\
BRX-TH-571\\
BOW-PH-135\\
NSF-KITP-05-93\\
\end{flushright}
\vspace{30mm}

\vspace*{.3in}

\begin{center}
{\Large\bf\sf  Level-rank duality of
D-branes on the SU($N$) group manifold}

\vskip 5mm Stephen G. Naculich\footnote{Research supported in part
by the NSF under grants PHY-0140281 and PHY-0456944, \\
\phantom{aaa}  
and grant PHY-9907949 through the KITP Scholars Program.}$^{,a}$
 and Howard J.  Schnitzer\footnote{Research supported in part 
by the DOE under grant DE--FG02--92ER40706.\\
{\tt \phantom{aaa} schnitzer@brandeis.edu; naculich@bowdoin.edu}\\
}$^{,b}$

\end{center}

\begin{center}
$^{a}${\em Department of Physics\\
Bowdoin College, Brunswick, ME 04011}

\vspace{.2in}

$^{b}${\em Martin Fisher School of Physics\\
Brandeis University, Waltham, MA 02454}
\end{center}
\vskip 2mm

\begin{abstract}
The consequences of level-rank duality for 
untwisted D-branes on an SU($N$) group manifold are explored.
Relations are found between the charges of D-branes 
(which are classified by twisted K-theory)
belonging to $\sunk$ and $\sukn$ WZW theories, 
in the case of odd $N+K$.
An isomorphism between the charge algebras 
is also demonstrated in this case.

\end{abstract}

\vfil\break

%\renewcommand{\baselinestretch}{2}
%\small\normalsize

\renewcommand{\theequation}{1.\arabic{equation}}
\setcounter{equation}{0}

\section{Introduction}

Understanding the nature of D-branes 
is a central issue of contemporary string theory,  
particularly the properties of D-branes 
in nontrivial gravitational and $B$-field backgrounds.  
(For a review, see ref.~\cite{Schomerus:2002dc}.)
One approach to this question is to study 
D-branes on group 
manifolds \cite{Klimcik:1996hp}-\cite{Fredenhagen:2004xp},
where the background is highly symmetric,
and the associated conformal field theory
(the WZW model) exactly solvable.
The D-branes in this theory correspond to
boundary states of the WZW model,
which can studied algebraically.
For the D-branes to be stable, 
the bosonic WZW theory should be regarded as 
part of a supersymmetric theory 
on the group manifold \cite{Fredenhagen:2000ei,Maldacena:2001xj};  
we will only consider the simplest case 
where the boson and fermion sectors are decoupled.
Strings on arbitrary group manifolds can also serve as 
building blocks of coset models.

Much can be learned about D-branes by studying their charges,
which are classified by K-theory or,
in the presence of a cohomologically nontrivial $H$-field background,
twisted K-theory \cite{Minasian:1997mm}.
The charge group for D-branes 
on a simply-connected group manifold $G$
with level $K$ is given by the twisted K-group
\cite{Fredenhagen:2000ei,Maldacena:2001xj,
Bouwknegt:2000qt,Freed:2001jd,Braun:2003rd,Gaberdiel:2002qa}
\be
\label{eq:Ktheory}
K^*(G) = \oplus_{i=1}^{m}
 \Z_x\,,\qquad
m= 2^{{\rm rank}\,G - 1}
\ee
where $\Z_x \equiv \Z/x\Z$ 
with $x$ an integer depending on $G$ and $K$. 
For $\sunk$, for example, $x$ is given 
by \cite{Fredenhagen:2000ei}
\be
\label{eq:xnk}
x_{N,K} \equiv {  N+K \over \gcd \{ N+K, \lcm \{ 1, \ldots, N-1\} \}  }\,.
\ee
One of the $\Z_x$ factors in the charge group corresponds 
to the charge of untwisted (symmetry-preserving) D-branes.
The remaining factors (for $N>2$) correspond to the charges
of other branes of the theory
\cite{Fredenhagen:2000ei,Maldacena:2001xj,Gaberdiel:2002qa}.

An intriguing aspect of WZW models is level-rank duality, 
a relationship between various quantities in 
the $\sunk$  model and corresponding quantities in the $\sukn$ model
\cite{Naculich:1990hg}--\cite{Altschuler:1989nm}.
Similar dualities occur for orthogonal and symplectic groups,
and also in Chern--Simons 
theories \cite{Naculich:1990pa}--\cite{Bourdeau:1991uu}.
Heretofore no analysis of level-rank duality
has been given for boundary WZW theories.  
In this paper, 
we begin the study of this issue by considering 
the relationship between untwisted D-branes 
on $\sunk$ and $\sukn$ group manifolds.  

In section 2, we review some necessary details of the WZW theory
without boundary, including simple current symmetries
and level-rank duality.
Section 3 contains a description of untwisted D-branes of the
$\sunk$ WZW theory with boundary;
these D-branes are labelled by 
irreducible representations $\lam$ of $\sun$ 
that correspond to integrable highest-weight representations
of $\sunk$.
In section 4,
we demonstrate the relationship between 
the charge $Q_\lam$ of an $\sunk$ D-brane 
and that of a related D-brane of $\sukn$.  
Specifically, we find
\be
\tQ_\tlam = (-1)^{r(\lam)}  Q_\lam  ~~~  \mod x,  ~~~\forodd
\ee
where 
$\tQ_\tlam$ is the charge of the $\sukn$ D-brane  
labelled by the representation $\tlam$ of $\suk$
obtained by transposing the Young tableau of $\lam$,
$r(\lam)$ is the number of boxes of the Young tableau $\lam$,
and $x = \min\{x_{N,K}, x_{K,N} \}$.
(A similar but more complicated relationship is expected
when $N+K$ is even.\footnote{{\it Note added}: 
see ref.~\cite{Naculich:2006mt}.}) 
We also show that the charge algebra of D-branes
is isomorphic to the charge algebra of dual D-branes
in the level-rank dual WZW model (for $N+K$ odd),
and that the energies of level-rank dual D-branes
are equal.
Concluding remarks constitute section 5.

\renewcommand{\theequation}{2.\arabic{equation}}
\setcounter{equation}{0}

\section{Bulk WZW Theory and level-rank duality}

Strings on group manifolds are described by
the Wess-Zumino-Witten conformal field theory.
In this section, we review some aspects of the 
WZW model, including simple currents
and level-rank duality,
that will be needed in subsequent sections
to understand D-branes on group manifolds.

The WZW model is a rational conformal field theory
whose chiral algebra (for both left- and right-movers)
is the (untwisted) affine Lie algebra $\hgk$
at level $K$ with Virasoro central charge $c ={K \dim \g}/(K+h^\vee)$,
where $\g$ is the finite Lie algebra associated with $\hgk$,
and $h^\vee$ the dual Coxeter number of $\g$.
The building blocks of the WZW model 
are integrable highest-weight representations $V_\lam \in \Pplus$ of $\hgk$, 
which are labelled by $\lam$,
the irreducible representation of $\g$
spanning the lowest-conformal-weight-subspace 
of the $\hgk$ representation.
Associated with the affine Lie algebra $\hgk$
is an extended Dynkin diagram,
which has one more dot 
than the Dynkin diagram of $\g$.
Correspondingly, the highest-weight representation $V_\lam$
possesses an extra Dynkin index 
\be
a_0 = K - \sum_{i=1}^n  m_i a_i \,,\qquad 
n = {\rm rank~} \g \,,
\ee
where $a_i$ are the Dynkin indices of $\lam$ 
and the integers $m_i$ are the components of the highest co-root.
(For $\sunk$, the affine Lie algebra with which we will be primarily concerned,
$m_i=1$ for all $i$.)
An {\it integrable} highest-weight representation is one satisfying 
$ a_0 \ge 0 $.

For purposes of discussing level-rank duality, 
it is useful to describe irreducible representations of $\g$
in terms of Young tableaux.
For example,
an irreducible representation of $\sun$ with Dynkin indices $a_i$ 
corresponds to a Young tableau with $N-1$ or fewer rows, 
with row lengths 
\be
\ell_i = \sum_{j=i}^{N-1} a_j \,, \qquad 
i=1, \ldots, N-1  \,.
\ee
Let $r(\lam) = \sum_i \ell_i$ denote the number of boxes
of the tableau.
Representations $\lam$ corresponding to integrable highest-weight 
representations $V_\lam$ of $\sunk$ obey $\ell_1 (\lam) \le K$,
\ie, their Young tableaux have $K$ or fewer columns.

The set of affine characters of the integrable highest-weight representations
\be
\chi_\lam (\tau) = \Tr_{V_\lam} \e^{2\pi i \tau(L_0 - c/24)}\,, \qquad 
\lam \in \Pplus
\ee
is closed under the modular transformation $\tau \rightarrow -1/\tau$,
the mixing being described by the modular transformation matrix $S_{\mu\nu}$. 

A primary field $\phi_\lam$ of the conformal field
theory is associated with 
each integrable highest-weight representation $V_\lam$.
The multiplicities of the primary fields appearing 
in the operator product expansion of 
a pair of primary fields of the WZW model are given by the fusion
coefficients $\Nabc$ appearing in the fusion algebra 
\be
\label{eq:fusionalgebra}
\phi_\mu \cdot \phi_\nu = \sum_{\lam \in \Pplus}  \Nabc ~\phi_\lam 
\ee
where $\Nabc$ is given by Verlinde's formula \cite{Verlinde:1988sn}
\be
\label{eq:verlinde}
\Nabc = \sum_{\rho \in \Pplus}  
{ S_{\mu\rho} S_{\nu\rho} S^*_{\lam\rho} \over S_{\id\rho} } ~,
\ee
with $\id$ denoting the identity representation.
For fixed $\mu$ and $\nu$,
and for $K$ sufficiently large,
the fusion coefficients $\Nabc$ become equal to  $\bNabc$,
the multiplicities appearing in the 
tensor product decomposition of representations of $\g$
 \be
\label{eq:LRcoeffs}
\mu \otimes \nu = \bigoplus_\lam  \bNabc ~\lam  ~.
\ee
In the case of $\sunk$,
``sufficiently large'' means $\ell_1(\mu)+\ell_1(\nu) \le K$,
and $\bNabc$ are just the Littlewood-Richardson coefficients.

In this paper, we only consider 
the WZW theory with a diagonal closed-string spectrum:
\be
\cH^{\rm closed}  
= \bigoplus_{\lam \in \Pplus}  
 V_\lam \otimes \bV_{\lam^*}
\ee
where $\bV$ represents right-moving states,
and $\lam^*$ denotes the representation conjugate to $\lam$,
\ie, such that ${N_{\lam \lam^*}}^\id = 1$.
The partition function for this theory
\be
Z(\tau) 
= \sum_{\lam \in \Pplus} \left|  \chi_\lam (\tau)  \right|^2
\ee
is automatically modular invariant.

\vs{.1in}
\noindent{\bf Simple current symmetries}
\vs{.1in}

\option
Automorphisms of the extended Dynkin diagram 
shuffle the Dynkin indices
and thus relate different integrable representations to one another.
For example, the extended Dynkin diagram of $\sunk$ 
has a $\Z_N$ symmetry $\sig$
which takes a representation $\lam$ into $ \lam' = \sig (\lam)$,
whose Dynkin indices are 
$ a'_i = a_{i-1} $ for $ i=1, \ldots, N-1,$ and $ a'_0 = a_{N-1}$.
The Young tableau for $\lam'$ is obtained by
placing a row of length $K$ on top of the tableau for $\lam$,
and deleting any columns of length $N$ that may result.  
The modular transformation matrix 
for $\sunk$ transforms under $\sig$ as \cite{Altschuler:1989nm}
\be
S_{\sig(\mu) \nu} = \e^{  -2\pi i r(\nu) / N } \Smat \,,
\ee
as a result of which the fusion rule coefficients 
for $\sunk$ satisfy
\be
\label{eq:sigmafusion}
  { N_{\sig^m(\mu) \sig^n(\nu)}}^{\sig^l(\lam)} 
        = \Nabc      {\rm ~~~~~for~} m+n=l  \mod N~ \,. 
\ee  
The orbits 
$\{ \lam$,  $\sig(\lam), \ldots$, $\sig^{N-1}(\lam) \}$ 
of integrable highest-weight representations
are termed ``cominimal equivalence classes.''
(Some orbits may have fewer elements.)
The members of the orbit of the identity representation
$\sig^j(\id)$, with $j = 1, \ldots, N-1$, 
are termed ``cominimal representations''
or ``simple currents,''
and correspond to rectangular Young tableaux 
with $j$ rows and $K$ columns.
The fusion algebra for these representations 
\be
\label{eq:simplecurrent}
\phi_\lam \cdot \phi_{\sig^j(\id)} =  \phi_{\sig^j(\lam)}
\ee
contains only one term by virtue of eq.~(\ref{eq:sigmafusion}).

\vs{.1in}
\noindent{\bf Level-rank duality}
\vs{.1in}

\option
An intriguing relation, level-rank duality,
exists between the WZW model for $\sunk$
and the corresponding WZW model with $N$ and $K$ 
exchanged \cite{Naculich:1990hg}-\cite{Bourdeau:1991uu}.
The Young tableau $\lam$ corresponding to an integrable highest-weight
representation of $\sunk$ maps under transposition
(\ie, exchange of rows and columns) to a Young tableau $\tlam$ 
that corresponds to an integrable highest-weight
representation of $\sukn$ 
(possibly after removing any columns of length $K$).
This map is not one-to-one,
since cominimally-equivalent representations of $\sunk$
may map into the same representation of $\sukn$
(due to the removal of columns).
It is clear, however, that the cominimal equivalence classes 
of the two theories {\it are} in one-to-one correspondence. 

The modular transformation matrices and fusion rule coefficients
of the $\sunk$ theory
obey simple relations under the exchange of $N$ and $K$.
Letting $\Smat$ and $\tSmat$ denote the modular transformation 
matrices of $\sunk$ and $\sukn$, one finds \cite{Altschuler:1989nm}
\be
\Smat = \sqrt {K\over N} \ \e^{-2\pi i  r(\mu) r(\nu)/NK} \ \tSmat^* ~.
\ee
{}From this and eq.~(\ref{eq:verlinde}), 
it follows that \cite{Altschuler:1989nm}
\be
\label{eq:fusionduality}
\Nabc  = \tNabc \,,
\qquad\qquad 
\Dr =  {r(\mu) + r(\nu) - r(\lam) \over N} \in \Z
\ee
where $\tN$ denotes the fusion rule multiplicities of $\sukn$.
(For $N$ sufficiently large ---
\viz, for $N > k_1 (\mu) + k_1 (\nu) $,
where $k_1(\mu)$ denotes the length of the first column  of $\mu$ ---
$\Dr$ vanishes,
so that on the right-hand side of the fusion algebra (\ref{eq:fusionalgebra}),
$\lam$ is simply dual to $\tlam$, its transpose,
but in general the relation is more complicated.)

\section{Boundary WZW Theory and D-brane charges}

\renewcommand{\theequation}{3.\arabic{equation}}
\setcounter{equation}{0}

D-branes on group manifolds have received a lot of attention,
from both the algebraic and geometric point of 
view \cite{Klimcik:1996hp}-\cite{Fredenhagen:2004xp}.
Algebraically, D-branes on group manifolds 
can be studied in terms of the possible boundary conditions 
that can imposed on a WZW model with boundary.
Let the open string world-sheet be the upper half plane.
The nonvanishing components of the stress-energy tensor 
must satisfy $T(z) = \bT(\bz)$ on the boundary $z=\bz$.
Additional restrictions may be imposed on the currents 
of the affine Lie algebra on the boundary, \eg
\be
\label{eq:boundaryconditions}
\left[ J^a(z) - \omega \bJ^a(\bz)\right] \bigg|_{z=\bz} = 0\,,
\ee
where $\omega$ is an automorphism of the affine Lie algebra,
although eq.~(\ref{eq:boundaryconditions}) 
is not required by the conformal symmetry.
Open-closed string duality 
correlates the boundary conditions (\ref{eq:boundaryconditions}) 
of the boundary WZW model 
with certain coherent states $\boundary$ of the bulk WZW model 
satisfying 
\be
\label{eq:modes}
\left[  J^a_n + \omega \bJ^a_{-n} \right] \boundary = 0
\ee
where $J^a_n$ are the modes of the current algebra generators.

Symmetry-preserving, or untwisted, D-branes correspond to $\omega=1$,
and it is this special class of branes that will be the focus of this paper.
For $\omega=1$, 
equation (\ref{eq:modes}) can be satisfied by a state 
belonging to a single sector $ V_\lam \otimes \bV_{\lam^*} $
of the WZW theory;
such states are termed ``Ishibashi states'' $\ishilam$ \cite{Ishibashi:1988kg}.
Because we are considering the diagonal WZW theory, 
all states $\ishilam$ for $\lam \in \Pplus$ 
belong to the spectrum of the bulk WZW theory.

Boundary states corresponding to D-branes must satisfy additional
(Cardy) conditions \cite{Cardy:1989ir}, 
whose solution may be written as certain linear combinations of 
the Ishibashi states known as Cardy states $\cardy$.
For untwisted D-branes of the diagonal WZW theory,
these states are of the form 
\be
\cardy = \sum_{\mu \in \Pplus}  
{S_{\lam \mu} \over \sqrt{S_{\id\mu}}} \ishimu \,.
\ee

{}From a geometric point of view
(and for large values of $K$),
the untwisted D-branes wrap conjugacy classes 
on the group manifold 
and are stabilized by flux stabilization 
\cite{Alekseev:1998mc}--\cite{Maldacena:2001xj}.
Quantization imposes constraints on the allowed conjugacy classes,
which are in one-to-one correspondence with integrable highest-weight 
representations $V_\lam \in \Pplus$ of $\hgk$.

\vs{.1in}
\noindent{\bf D-brane charges}
\vs{.1in}

\option
Next we consider the charge of the untwisted D-brane 
associated with the boundary state $\cardy$.
The state $\idcardy$ corresponds geometrically to a D0-brane 
located at the identity element of the group manifold,
to which we assign unit charge, $Q_\id=1$.
A collection of $n$ such D0-branes has D0-charge $n$.
By renormalization group flow arguments presented in 
refs.~\cite{Affleck:1990by,Alekseev:1999bs,
Bachas:2000ik,Alekseev:2000jx,Fredenhagen:2000ei,Maldacena:2001xj},
these D0-branes may form a bound state (D-brane) 
associated with the Cardy state $\cardy$, 
corresponding to an $n$-dimensional representation $\lam$ of $\g$.
Hence the D0-charge $Q_\lam $ of this D-brane is $(\dim \lam)$,
but is only defined modulo some 
integer $x$ \cite{
Alekseev:2000jx,Fredenhagen:2000ei,Maldacena:2001xj,Bouwknegt:2002bq}.

By considering condensation of D-branes,
one finds that the charges must obey the fusion 
algebra \cite{Fredenhagen:2000ei}
\be
\label{eq:chargealgebra}
Q_\mu \cdot Q_\nu = \sum_{\lam \in \Pplus}  \Nabc ~Q_\lam \,.
\ee
{}From eq.~(\ref{eq:LRcoeffs}), one has 
\be
(\dim \mu) (\dim \nu) = \sum_{\lam}  \bNabc (\dim \lam)
                      \ge \sum_{\lam \in \Pplus}  \Nabc (\dim \lam)\,.
\ee
For sufficiently large $K$, 
the last inequality is saturated,
in which case eq.~(\ref{eq:chargealgebra}) is consistent with 
$Q_\lam = \dim \lam$.
In general, however, the charge algebra (\ref{eq:chargealgebra}) is
only satisfied modulo $x$,
which is the largest integer for which 
\be
(\dim \mu) (\dim \nu) = \sum_{\lam \in \Pplus}  \Nabc (\dim \lam) 
\qquad \mod~x
\ee
holds for all $\mu$, $\nu \in \Pplus$.

Now we turn to the specific case of $\sunk$.
To determine the value of $x$, 
it is sufficient \cite{Fredenhagen:2000ei} to consider the fusion algebra 
(\ref{eq:simplecurrent}) 
involving simple currents and the 
fundamental representations $\Lam_s$ of $\sun$,
$s=1, \ldots, N-1$, whose Young tableaux  are 
$ \oneoneoneone \} s $. 
This implies
\bea
\label{eq:simplecurrentfusion}
(\dim \Lam_s) (\dim \sig(\id) ) 
&=&  
\dim {\sig(\Lam_s)} \qquad \mod~x \, ,
\nonumber\\ [.1in]
\left(   N \atop s \right) \left(  N+K-1 \atop K \right) 
&=& 	
{ (N+K-1)! \over (N-s)! (K-1)! s! (K+s) } 
\qquad \mod~x  \,.
\eea
Then $x$ is given by greatest common denominator of the 
difference between the two sides of eq.~(\ref{eq:simplecurrentfusion})
\be
\label{eq:diff}
x = \gcd 
\left\{  
	{s \over K+s}  \left( N+K \atop K \right) \left( N \atop s \right)
           \bigg| s=1, \ldots, N-1 
\right\}  \,.
\ee
In refs.~\cite{Fredenhagen:2000ei,Bouwknegt:2002bq},
it is shown that eq.~(\ref{eq:diff}) implies $x=x_{N,K}$  where 
\be
x_{N,K} \equiv {  N+K \over \gcd \{ N+K, \lcm \{ 1, \ldots, N-1\} \}  } \,.
\ee
Hence, the charge algebra (\ref{eq:chargealgebra})
is satisfied provided the charges of the untwisted D-branes in $\sunk$ 
are defined modulo $x_{N,K}$,
\be
 Q_\lam = \dim \lam \;\;\; \mod   \: x_{N,K} \qquad {\rm for}~~ \sunk\,.
\ee
(The values of $x$ for all other simple groups is given 
in ref.~\cite{Bouwknegt:2002bq}.)
Thus untwisted D-branes correspond to the first factor 
$\Z_{x_{N,K}}$ of the twisted K-theory group (\ref{eq:Ktheory}).

Finally, we consider the relation between the charges of $\sunk$
D-branes 
corresponding to the cominimally-equivalent Cardy states
$\cardy$ and $\sigcardy$.
The action of $\sigma$ on integrable highest-weight representations
corresponds geometrically to the action of the center of $\g$ on the 
conjugacy classes of $\g$ \cite{Alekseev:2000jx,Maldacena:2001xj}.
The rotation of a conjugacy classes, however, cannot change the
(magnitude of the) charge of the associated brane, 
thus \cite{Maldacena:2001xj}
\be
\label{eq:sigmacharge}
Q_{\sig (\lam)} = (-1)^{N-1}  Q_\lam \quad\mod x_{N,K}
\qquad {\rm for}~~ \sunk 
\ee
where the relative sign comes from the action of $N-1$ elements 
of the Weyl group,
each reflection changing the orientation of the D-brane.

\section{Level-rank duality of WZW D-branes}

\renewcommand{\theequation}{4.\arabic{equation}}
\setcounter{equation}{0}

In this section, we will establish a relation between the charges of 
untwisted 
D-branes of $\sunk$ and those of the level-rank dual theory $\sukn$.

\vs{.1in}
\noindent{\bf Level-rank duality of D-brane charges} 
\vs{.1in}

\option
Level-rank duality relates the cominimal equivalence classes
of $\sunk$ to  those of $\sukn$.
Since untwisted D-branes are labelled by integrable highest-weight 
representations,
and since by eq.~(\ref{eq:sigmacharge})
their charges are invariant (modulo sign and modulo $x$) under
the operation $\sigma$, and therefore depend only on the
cominimal equivalence class of the representation, 
it would be reasonable to expect 
the charges of level-rank dual D-branes to be equal 
(modulo sign and modulo $x$).
Indeed we will demonstrate below that this expectation is
borne out, provided $N+K$ is odd.
The situation is unclear for $N+K$ even.

Since the charges of $\sunk$ D-branes are only defined modulo $x_{N,K}$,
and the charges of $\sukn$ D-branes are only defined modulo $x_{K,N}$,
comparisons of these charges is only possible 
modulo $ \gcd \{ x_{N,K} ,x_{K, N} \} $.
For the remainder of this section, we assume
without loss of generality that $N \ge K$,
in which case $ \gcd \{x_{N,K}, x_{K, N} \} = x_{N,K} $.
(Thus the group with the larger rank has the smaller periodicity.)
All D-brane charges will be considered modulo $x \equiv x_{N,K}$.

The relation (which we prove below) between $Q_\lam$, 
the charge of the D-brane corresponding to the state $\cardy$ of $\sunk$,
and $\tQ_\tlam$, 
the charge of the level-rank dual D-brane $\dualcardy$ of $\sukn$,
is
\be
\label{eq:chargeduality}
\tQ_\tlam = (-1)^{r(\lam)}  Q_\lam  ~~~  \mod x,  ~~~\forodd\,.
\ee
To establish eq.~(\ref{eq:chargeduality}), we need to prove
\be
\label{eq:dimmodx}
(\dim \tlam)_\suk = (-1)^{r(\lam)}  (\dim \lam)_\sun \quad 
\mod x,   \qquad\forodd
\ee
for any integrable highest-weight representation of $\sunk$.

\vs{.1in}
\noindent {\it Proof:} 
Consider the branching of a representation $\lam$ of $\sump$ 
into representations $(\lam', \lam'')$ of $\sumo \times \sumt$
$$
(\dim \lam)_{\sump} 
= \sum_{\lam', \lam''} \LR
(\dim \; \lam')_{\sumo} (\dim \; \lam'')_{\sumt} 
 $$
where the integers $\LR$ 
are the Littlewood-Richardson coefficients \cite{FultonHarris}.
Formally, we let $M_1 = N+K$ and $M_2 = -K$,
and then use
\be
(\dim \; \lam'')_{\suminusk} =  (-1)^{r(\lam'')} (\dim \; \tlam'')_{\suk}  
\ee
to obtain
\be
\label{eq:decomp}
\!\!\!\!\!
(\dim \lam)_{\sun}
= (-1)^{r(\lam)} (\dim \tlam)_{\suk}
+ \sum_{\lam'\neq\id \atop \lam''} (-1)^{r(\lam'')} \LR
(\dim \; \lam')_{\sup} (\dim \; \tlam'')_{\suk} 
\ee
where we have separated the $\lam'=\id$ term from the rest of the sum.

First consider the case where $N+K$ is prime.
Using the hook length formula for the dimension of a representation
\be
(\dim \; \lam')_{\sup} = \prod_{(i,j)} \frac{ N+K+j-i}{h_{ij}} \,,
\ee
where the product is over the boxes $(i,j)$ of the Young tableau,
labeled by their row $i$ and column $j$,
and $h_{ij}$ are the corresponding hook lengths, 
we see that, for all $\lam' \neq \id$, 
the numerator contains the factor $N+K$.
The maximum hook length $h_{11}$ 
is given by $\ell_1 + k_1 -1$,
where $\ell_i$ and $k_i$ are the row and column lengths
respectively of the Young tableau.
Since $\lam$ is an integrable representation of $\sunk$, 
its maximum hook length is $N+K-2$, 
and therefore this is also true for $\lam'$.
Since none of the hook lengths divide $N+K$ (prime),
we have that $ (\dim \; \lam')_{\sup} $
is a multiple of $N+K$ for all $\lam' \neq \id$,
thus the sum in eq.~(\ref{eq:decomp}) 
is divisible by $N+K$, 
which establishes eq.~(\ref{eq:dimmodx}) when $N+K$ is prime,
since $x=N+K$ in this case.

For $N+K$ not prime, we must use a different approach.
The dimension of an arbitrary irreducible representation 
of $\sun$ can be written as the determinant of an 
$\ell_1 \times \ell_1$ matrix \cite{FultonHarris}
\be
\label{eq:Giambelli}
\dim \lam = \Big| \dim \Lam_{k_i+j-i} \Big| \;, 
\qquad i,j = 1, \ldots, \ell_1
\ee
where $\Lam_s$ is the completely antisymmetric representation of 
$\sun$, whose Young tableau is $ \oneoneoneone \} s $.
The maximum value of $s$ appearing in eq.~(\ref{eq:Giambelli})
is $k_1 + \ell_1 - 1$, which is bounded by $N+K-2$
for integrable highest-weight representations of $\sunk$.
For $1\le s \le N-1$,  
$\Lam_s$ are the fundamental representations of $\sun$,
with $\dim \Lam_s = \left( N \atop s \right)$.
The representations $\Lam_0$ and $\Lam_N$ 
both correspond to the identity representation with dimension 1.
We define $\dim \Lam_s = 0$ for $s<0$ and for $s>N$.

There is an alternative formula for the dimension of 
a representation in terms of the determinant of a $k_1 \times k_1$ matrix, 
\viz, \cite{FultonHarris}
\be
\label{eq:Giambellitwo}
\dim \lam = \left| \dim \tLam_{\ell_i+j-i} \right| \;, \
\qquad i,j = 1, \ldots, k_1
\ee
where $\tLam_s$ is the completely symmetric representation of $\sun$,
whose Young tableau is $\underbrace{\four}_s$.
For $s \ge 0$, $\dim \tLam_s = \left(N +s-1 \atop s \right)$.
We define $\dim \tLam_s = 0 $ for $s<0$.

To establish level-rank duality of D-brane charges using
eqs.~(\ref{eq:Giambelli}) and (\ref{eq:Giambellitwo}), 
we need to obtain the relation between 
$(\dim\Lam_s)_\sun$ and $(\dim\tLam_s)_\suk$ 
for all $s \le N+K-2$.
We consider three cases separately:
\begin{itemize}
\item $s \le N-1$:

For the fundamental representations of $\sun$,
we use eq.~(\ref{eq:decomp}) to establish that
\be
\label{eq:levelrankfund}
(\dim \Lam_s)_{\sun} = (-1)^{s} (\dim \tLam_s)_{\suk} \quad \mod x, 
\qquad s \le N-1
\ee
since each term in the sum on the r.h.s.~of eq.~(\ref{eq:decomp})
includes a factor 
$(\dim \; \Lam_t)_{\sup} = \left( N+K \atop t \right)$, 
$1 \le t \le s$,
and 
\be
\label{eq:appC}
\gcd   \left\{ \left. \left(   N+K \atop t \right)  \right|  
t = 1, \ldots, N-1 \right\} =  x \,,
\ee
as shown in Appendix C of ref.~\cite{Maldacena:2001xj}. 
Equation (\ref{eq:levelrankfund}) also applies trivially when $s \le 0$.

\item $N+1 \le s \le N+K-2$:

For this case, $(\dim \Lam_s)_\sun$ vanishes,
so eq.~(\ref{eq:levelrankfund}) will hold provided that
$ (\dim \tLam_s)_{\suk} $ vanishes mod $x$.
To show this, we repeatedly use the identity 
$ \left( M+1 \atop j \right) 
= \left( M \atop j \right)
+ \left( M \atop j-1 \right)$
to show that
\be
\label{eq:secondcase}
(\dim \tLam_s)_{\suk}  =  \left( K+s-1 \atop s \right)
= \sum_{u=0}^{s-N-1}  \left(s-N-1 \atop u\right) \left(N+K \atop N+u+1 \right),
\quad N+1 \le s \le N+K-2
\ee
Since $ \left(N+K \atop N+u+1 \right) =  \left(N+K \atop K-u-1 \right)$,
and recalling that $K \le N$,
we see that each term in the sum in eq.~(\ref{eq:secondcase}) 
contains a factor 
belonging to the set in eq.~(\ref{eq:appC}),
and therefore 
$(\dim \tLam_s)_{\suk} $ is a multiple of $x$.

\item $s=N$:

The remaining case is easily evaluated using eq.~(\ref{eq:sigmacharge}):
\bea
(\dim \Lam_N)_{\sun} &=& (\dim \id)_{\sun} = (\dim \id)_{\suk} 
\nonumber\\
&=&  (-1)^{K-1} (\dim \sig(\id))_{\suk}  \quad \mod x
\nonumber\\
&=& (-1)^{N-K-1} \left[ (-1)^N (\dim \tLam_N)_{\suk} \right] \quad \mod x
\eea
which is in accord with eq.~(\ref{eq:levelrankfund}),
but {\it only} when $N+K$ is odd.
\end{itemize}

\noindent To summarize, we have shown that 
\be
\label{eq:allanti}
(\dim \Lam_s)_{\sun} = (-1)^{s} (\dim \tLam_s)_{\suk} \quad \mod x, 
\quad
s \le N + K - 2, \quad \forodd \,.
\ee
The equality (\ref{eq:allanti}) also holds for $N+K$ even, except
when $s=N$, in which case the sign is reversed.

Restricting ourselves to $N+K$ odd,
we use eq.~(\ref{eq:allanti}) in eq.~(\ref{eq:Giambelli})
to find
\bea
(\dim \lam)_\sun  
&=& \left| (-1)^{k_i+j-i} (\dim \tLam_{k_i+j-i})_\suk \right| \quad \mod x
\nonumber\\
&=& (-1)^{r(\lam)}
\left| (\dim \tLam_{k_i+j-i})_\suk  \right| \quad \mod x, 
\qquad\forodd
\eea
where $r(\lam) = \sum_{i=1}^{\ell_1} k_i(\lam)$.
Comparing this with eq.~(\ref{eq:Giambellitwo}),
we see that (provided $\ell_1 < K$)
the r.h.s.~is the dimension of a representation
with row lengths $k_i$ and column lengths $\ell_i$, 
that is, the transpose representation $\tlam$,
hence
\be
\label{eq:dimduality}
(\dim \lam)_\sun  = (-1)^{r(\lam)} (\dim \tlam)_\suk \;\; \mod x,
\qquad \forodd
\ee
from which follows the level-rank duality of D-brane charges
(\ref{eq:chargeduality}).  
One subtlety remains:  if $\ell_1 = K$ for $\lam$, 
then the transpose $\tlam$ contains leading columns of $K$ boxes.
In that case, one can use eq.~(\ref{eq:sigmacharge}) $k_K$ times
to relate $\lam$ to a tableau $\hlam$ with no rows of length $K$,
and then apply eq.~(\ref{eq:dimduality}).
The overall prefactor is then $(-1)^{r(\hlam) + (N-1) k_K}$,
which is equal to $(-1)^{r(\hlam) + K k_K} = (-1)^{r(\lam)}$ 
since $N+K$ is odd, so that eq.~(\ref{eq:dimduality}) holds
in this case as well.
{\it QED.}
\vs{.1in}

The failure of eq.~(\ref{eq:allanti})
to hold for $s=N$ when $N+K$ is even, however,
precludes (\ref{eq:dimduality}) from holding generally in this case.
The adjoint of SU$(3)_3$ provides a simple counterexample.
If, however,
the maximum hook length of $\lam$ (\viz, $\ell_1 + k_1 -1$)
is less than $N$, then eq.~(\ref{eq:dimduality}) also holds for 
$N+K$ even.

At present,\footnote{See footnote 3.}
we do not know the precise relation between
the charges of level-rank dual D-branes for even $N+K$ 
when the maximum hook length is equal to or greater than $N$. 
(In many cases where $N+K$ is even and $N \neq K$, however, 
$x$ is unity, so eq.~(\ref{eq:chargeduality}) is trivially satisfied.)

\vs{.1in}
\noindent{\bf Level-rank duality of the charge algebra}
\vs{.1in}

\option
The level-rank duality of D-brane charges proved above,
together with the previously-known level-rank duality for
fusion coefficients, 
can be used to show that the charge algebras
associated with $\sunk$ and $\sukn$ are 
isomorphic, in the case that $N+K$ is odd.

We begin with the charge algebra for $\sunk$
\be
\label{eq:charge}
Q_\mu \cdot Q_\nu = \sum_{\lam \in \Pplus}  \Nabc ~Q_\lam 
\ee
or
\be
(\dim \mu)_\sun (\dim \nu)_\sun 
= \sum_{\lam \in \Pplus}  \Nabc ~(\dim \lam)_\sun \quad \mod~x \,.
\ee
Using eq.~(\ref{eq:dimduality}), we have
\be
(\dim \tmu)_\suk (\dim \tnu)_\suk 
= \sum_{\lam \in \Pplus}  (-1)^{N \Delta}
\Nabc ~(\dim \tlam)_\suk  ~~~\mod~x, \qquad \forodd
\ee
where $\Delta = \left[ r(\mu) + r(\nu) - r(\lam) \right]/N \in \Z$.
Since $N+K$ is odd, 
we have $(-1)^{N \Delta} = (-1)^{(K-1)\Delta}$, 
which together with eq.~(\ref{eq:sigmacharge}) yields
\be
(\dim \tmu)_\suk (\dim \tnu)_\suk 
= \sum_{\lam \in \Pplus}  
\Nabc ~(\dim \sig^\Dr (\tlam) )_\suk  ~\mod~x, \qquad \forodd \,.
\ee
Finally,
we use the level-rank duality of fusion coefficients (\ref{eq:fusionduality})
to obtain the level-rank dual charge algebra
\be
(\dim \tmu)_\suk (\dim \tnu)_\suk 
= \sum_{\lam \in \Pplus}  
\tNabc ~(\dim \sig^\Dr (\tlam) )_\suk  ~\mod~x, \qquad \forodd \, .
\ee
That the D-brane charges  of $\sukn$ satisfy the $\sukn$ charge-algebra
\be
\label{eq:dualcharge}
\tQ_\tmu \cdot \tQ_\tnu = \sum_{\lam \in \Pplus}  \tNabc ~\tQ_{\sig^\Dr(\tlam)} 
\ee
is obvious, 
but what we have shown is that 
the solutions to the $\sunk$ charge algebra
are isomorphic (modulo $x$) to 
the solutions to the charge algebra of the dual D-branes in the 
$\sukn$ theory, when $N+K$ is odd.
That is, term-by-term, the right hand sides 
of eqs.~(\ref{eq:charge}) and (\ref{eq:dualcharge}) match:
each term $Q_\lam$ is equal (mod $x$) to the 
corresponding term $\tQ_{\sig^\Dr(\tlam)}$,
possibly modulo a sign common to all the terms in the sum.

Note that this isomorphism does not hold for $N+K$ even.
A simple counterexample is  
$\one \otimes \one = 1 \oplus \two $ in SU$(2)_2$,
where $\dim(\two)=3$ mod 4 is not equal to $\dim(\oneone)=1$ mod 4.

\vs{.1in}
\noindent{\bf Level-rank duality of the masses of Cardy states}
\vs{.1in}

\option
In ref.~\cite{Maldacena:2001xj}, 
it is shown that the mass of a boundary state, 
normalized to that of the identity representation,  
is given by
\be
\frac{ {\rm Energy} (\cardy)}
      { {\rm Energy} (\idcardy)}
  =  \frac{S_{\id\lam}}{S_{\id\id}}
\ee
where the r.h.s.~of this equation is simply the $q$-dimension of $\lam$
where $q = \e^{2\pi i/(N+K)}$.
It was shown in refs.~\cite{Naculich:1990pa}
that $q$-dimensions of primary fields are invariant
under level-rank duality 
\be
\left( {S_{\id\lam} \over S_{\id\id} } \right)_{\sunk}
= 
\left(\frac{\tS_{\id\tlam}}{\tS_{\id\id}} \right)_{\sukn}
\ee
hence the masses of the level-rank dual Cardy states are equal:
\be
\left( {\rm Energy} (\cardy) \over {\rm Energy} (\idcardy) \right)_{\sunk} 
=
\left( {\rm Energy} (\dualcardy) \over 
     {\rm Energy} (\idcardy) \right)_{\sukn} \,.
\ee

\renewcommand{\theequation}{5.\arabic{equation}}
\setcounter{equation}{0}

\section{Conclusions}

In this paper, we have begun to analyze the consequences of
level-rank duality in boundary WZW models.
We have found that the D0-charge of a symmetry-preserving
D-brane $\cardy$ of the $\sunk$ model is equal 
(up to a sign $(-1)^{r(\lam)}$, where $r(\lam)$ is the 
number of boxes of the Young tableau of $\lam$) 
to the charge of the level-rank dual D-brane $\dualcardy$ 
of the $\sukn$ model,
provided that $N+K$ is odd.
(A similar relation for even $N+K$,
but with a more complicated expression for the relative sign, 
is anticipated.\footnote{See footnote 3.})
The charges of D-branes are only defined modulo $x_{N,K}$,
given by eq.~(\ref{eq:xnk}),
which is related to the twisted K-theory group of $\sunk$.
Since the periodicities $x_{N,K}$ are not level-rank dual,
the charges of level-rank dual D-branes can only 
be compared modulo the periodicity of the smaller charge group.

We have also shown that the charge algebras of D-branes of 
level-rank dual theories are isomorphic, 
again when $N+K$ is odd. 
Finally, we observed that the masses of the level-rank
dual D-branes are equal,
which follows from the level-rank duality of quantum dimensions
of integrable highest weight representations.

\section*{Acknowledgments}

SN wishes to thank T. Pietraho for helpful remarks,
and the Kavli Institute for Theoretical Physics 
for hospitality in August 2005, 
when some of this work was done, 
and for support through the KITP Scholars program.
The authors also thank the referee for 
pointing out a lacuna in the proof in section 4.

\providecommand{\href}[2]{#2}\begingroup\raggedright\endgroup

\end{document}